\documentclass[aps,prb,twocolumn,showpacs,preprintnumbers,amsmath,amssymb,superscriptaddress]{revtex4}

\usepackage{graphicx}
\usepackage{dcolumn}
\usepackage{bm}

\newcommand{\bra}[1]{\langle #1|}
\newcommand{\ket}[1]{|#1\rangle}

\begin{document}

\title{Spin-polarized transport through weakly coupled double
quantum dots in the Coulomb-blockade regime}

\author{Ireneusz Weymann}
\email{weymann@amu.edu.pl} \affiliation{Department of Physics,
Adam Mickiewicz University, 61-614 Pozna\'n, Poland}

\date{\today}

\begin{abstract}
We analyze cotunneling transport through two quantum dots in
series weakly coupled to external ferromagnetic leads. In the
Coulomb blockade regime the electric current flows due to
third-order tunneling, while the second-order single-barrier
processes have indirect impact on the current by changing the
occupation probabilities of the double dot system. We predict a
zero-bias maximum in the differential conductance, whose magnitude
is conditioned by the value of the inter-dot Coulomb interaction.
This maximum is present in both magnetic configurations of the
system and results from asymmetry in cotunneling through different
virtual states. Furthermore, we show that tunnel magnetoresistance
exhibits a distinctively different behavior depending on
temperature, being rather independent of the value of inter-dot
correlation. Moreover, we find negative TMR in some range of the
bias voltage.
\end{abstract}


\maketitle

\section{Introduction}

Transport properties of double quantum dots (DQDs) have recently
attracted much interest.
\cite{ono02,derWielRMP03,hazalzetPRB01,golovachPRB04,graberPRB06,
cotaPRL05,mcclurePRL07,liuPRB05,wunschPRB06,franssonPRB06,franssonNT06,inarrea06,datta07}
This is because these structures are ideal systems to study the
fundamental interactions between electrons and spins.
\cite{wolf01,loss02,maekawa02,zutic04,maekawa06} Double quantum
dots exhibit a variety of interesting effects, such as for example
current rectification due to the Pauli spin blockade,
\cite{ono02,franssonPRB06,franssonNT06,inarrea06,datta07} negative
differential conductance, \cite{liuPRB05} formation of molecular
states, \cite{graberPRB06} spin pumping,
\cite{hazalzetPRB01,cotaPRL05} etc. Furthermore, double quantum
dots are also being considered for future applications in quantum
computing. \cite{lossPRA98} Experimentally, such systems may be
realized for example in lateral and vertical semiconductor quantum
dots. \cite{ono02,mcclurePRL07,liuPRB05,johnsonPRB05} Another
implementation of DQDs are single wall metallic carbon nanotubes
with top gate electrodes, which enable changing of charge on each
dot separately, as well as the intrinsic DQD parameters.
\cite{graberPRB06,jorgensenAPL06,sapmazNL06,graber06}

The goal of this paper is to analyze transport properties of DQDs
weakly coupled to ferromagnetic leads in the Coulomb blockade
regime. When the leads are ferromagnetic, transport strongly
depends on the magnetic configuration of the system, giving rise
to tunnel magnetoresistance (TMR), spin accumulation, exchange
field, etc. \cite{julliere75,barnas98,takahashi98,bulka00,
rudzinski01,koenigPRL03,braunPRB04,weymannPRB05,weymannPRB07} In
the Coulomb blockade regime the electric current flows due to
higher-order tunneling processes (cotunneling), while the
first-order tunneling processes (sequential tunneling) are
exponentially suppressed.
\cite{weymannPRB05,nazarov,kang,averin92} The problem of
spin-polarized cotunneling has been so far addressed mainly in the
case of single quantum dots.
\cite{weymannPRB05,weymannPRB07,weymannPRBBR05,braigPRB05,
weymannEPJ05,weymannPRB06,weymannEPL06,souza06} For example, it
was shown that tunnel magnetoresistance exhibits distinctively
different behavior depending on the number of electrons on the
dot. \cite{weymannPRB05} Moreover, the zero-bias anomaly was found
in the differential conductance when magnetic moments of the leads
form antiparallel configuration. \cite{weymannPRBBR05} Another
interesting behavior was predicted for quantum dots coupled to
ferromagnetic leads with non-collinear alignment of magnetizations
-- the exchange field was found to increase the differential
conductance for certain non-collinear configurations, as compared
to the parallel one. \cite{weymannPRB07,franssonEPL05} On the
other hand, it was shown experimentally for nonmagnetic systems
that the conductance of quantum dots in the cotunneling regime may
serve as a handle to determine the spectroscopic g-factor.
\cite{kogan04}

In the case of double quantum dots considered in this paper, in
the Coulomb blockade regime the electric current flows due to
third-order tunneling processes, while the single-barrier
second-order processes together with third-order processes
determine the double dot occupation probabilities. Assuming that
double quantum dot is occupied by two electrons in equilibrium,
one on each dot, we calculate the differential conductance $G$ and
tunnel magnetoresistance TMR. We show that differential
conductance exhibits a maximum at the zero bias. We further
distinguish two different mechanisms leading to this new zero-bias
anomaly. The first one is an asymmetry in cotunneling through
different virtual states of the DQD system, which leads to an
enhancement of $G$ at zero bias. Such asymmetry is induced by a
finite value of the inter-dot Coulomb interaction. This mechanism
is rather independent of magnetic configuration of the system. The
second mechanism leading to the zero-bias maximum in differential
conductance is the interplay between spin accumulation and
third-order tunneling processes carrying the current. This
mechanism does depend on the magnetic configuration of the system
and, as we show in the sequel, is found to be more efficient in
the antiparallel configuration. We also analyze the behavior of
TMR and show that the TMR exhibits a maximum at zero bias, which
strongly depends on the temperature. Furthermore, the TMR may
become negative in some range of the bias voltage.

Finally, we note that there are several experimental realizations
of single quantum dots attached to ferromagnetic leads,
\cite{chye02,ralph02,heersche06,zhang05,tsukagoshi99,
zhao02,jensen05,sahoo05,pasupathy04,fertAPL06,hamaya06} while
experimental data on spin-polarized transport through double
quantum dots is lacking. We believe that the results presented in
this paper will be of assistance in discussing future experiments.

\section{Model and method}

\begin{figure}[t]
  \includegraphics[width=0.9\columnwidth]{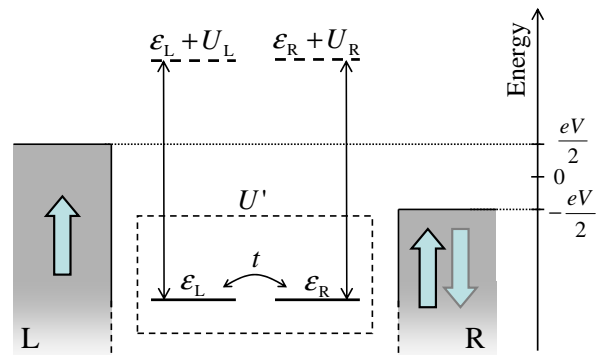}
  \caption{\label{Fig:1} (color online)
  Schematic of a double quantum dot coupled to ferromagnetic leads.
  The magnetic moments of the leads can form either
  parallel or antiparallel configuration.
  The system is symmetrically biased.}
\end{figure}

The schematic of a double quantum dot coupled to ferromagnetic
leads is shown in Fig.~\ref{Fig:1}. It is assumed that the
magnetizations of the leads are oriented collinearly, so that the
system can be either in the parallel or antiparallel magnetic
configuration. The Hamiltonian $\hat{H}$ of the DQD system is
given by, $\hat{H}=\hat{H}_{\rm L} + \hat{H}_{\rm R} +
\hat{H}_{\rm DQD} + \hat{H}_{\rm T}$. The first two terms describe
noninteracting itinerant electrons in the leads,
$\hat{H}_j=\sum_{{\mathbf k}\sigma} \varepsilon_{j{\mathbf
k}\sigma} c^{\dagger}_{j{\mathbf k}\sigma} c_{j{\mathbf k}\sigma}$
for the left ($j={\rm L}$) and right ($j={\rm R}$) lead, where
$\varepsilon_{j{\mathbf k}\sigma}$ is the energy of an electron
with the wave vector ${\mathbf k}$ and spin $\sigma$ in the lead
$j$, and $c^{\dagger}_{j{\mathbf k}\sigma}$ ($c_{j{\mathbf
k}\sigma}$) denotes the respective creation (annihilation)
operator. The double dot is described by the Hamiltonian
\begin{eqnarray}\label{Eq:DQDHamiltonian}
  \hat{H}_{\rm DQD} &=&
  \sum_{j={\rm L,R}}\sum_{\sigma}\varepsilon_{j} n_{j\sigma}
  + \sum_{j={\rm L,R}} U_{j} n_{j\uparrow}n_{j\downarrow}
  \nonumber\\
  &+& U^\prime \sum_{\sigma\sigma^\prime}
  n_{\rm L\sigma}n_{\rm R\sigma^\prime}
 \,,
\end{eqnarray}
with $n_{j\sigma}=d^{\dagger}_{j\sigma}d_{j\sigma}$, where
$d^{\dagger}_{j\sigma}$ ($d_{j\sigma}$) is the creation
(annihilation) operator of an electron with spin $\sigma$ in the
left ($j={\rm L}$) or right ($j={\rm R}$) quantum dot, and
$\varepsilon_{j}$ is the corresponding single-particle energy. The
Coulomb interaction on the left (right) dot is described by
$U_{\rm L}$ ($U_{\rm R}$). The last part of $\hat{H}_{\rm DQD}$
corresponds to the inter-dot Coulomb correlation, whose strength
is given by $U^\prime$. As we are interested in the low bias
voltage regime where the system is in the Coulomb blockade, it is
justifiable to assume that the energy level of each dot is
independent of the bias voltage. For the sake of clarity of
further discussion we also assume $\varepsilon_{\rm
L}=\varepsilon_{\rm R} \equiv \varepsilon$ and $U_{\rm L} = U_{\rm
R} \equiv U$.

We note that in a general case, the exchange interaction between
spins in the two dots may lead to the formation of singlet and
triplet states. \cite{graberPRB06} However, this exchange
interaction was found to be rather small as compared to the other
energy scales, \cite{ono02} and thus, following previous
theoretical works, \cite{wunschPRB06,aghassi06} we will neglect
it.

Tunneling processes between the two dots and electrodes are
described by the Hamiltonian,
\begin{eqnarray}
  \hat{H}_{\rm T} = \sum_{j={\rm L,R}}\sum_{{\mathbf k}\sigma}\left(t_{j}
  c^{\dagger}_{j {\mathbf k}\sigma}
  d_{j \sigma} + h.c.\right)
  + \left(t d^\dagger_{\rm L\sigma} d_{\rm R\sigma}
  + h.c. \right)
   \,,
\end{eqnarray}
where $t_j$ denotes the tunnel matrix elements between the $j$th
lead and the $j$th dot, and $t$ describes the hopping between the
two quantum dots. Coupling of the $j$th dot to the $j$th lead can
be expressed as $\Gamma_{j}^{\sigma}= 2\pi |t_{j}|^2
\rho_j^\sigma$, with $\rho_j^\sigma$ being the spin-dependent
density of states of the corresponding lead. With the definition
of the spin polarization of lead $j$, $p_{j}=(\rho_{j}^{+}-
\rho_{j}^{-})/ (\rho_{j}^{+}+ \rho_{j}^{-})$, the coupling can be
expressed as, $\Gamma_{j}^{+(-)}=\Gamma_{j}(1\pm p_{j})$, with
$\Gamma_{j}= (\Gamma_{j}^{+} +\Gamma_{j}^{-})/2$. Here,
$\Gamma_{j}^{+}$ and $\Gamma_{j}^{-}$ describe the coupling of the
$j$th dot to the spin-majority and spin-minority electron bands of
lead $j$, respectively. As reported in Ref.~\onlinecite{kogan04},
typical values of the coupling strength are of the order of tens
of $\mu$eV. In the following, we assume symmetric couplings,
$\Gamma_{\rm L} = \Gamma_{\rm R} \equiv \Gamma/2$, and equal spin
polarizations of the leads, $p_{\rm L} = p_{\rm R} \equiv p$.

In this paper we analyze spin-dependent transport through double
quantum dot in the case of the Coulomb blockade regime. We assume
that in equilibrium each dot is singly occupied, so that there are
two electrons in the DQD system. This transport regime can be
realized for example in lateral quantum dots
\cite{mcclurePRL07,liuPRB05,johnsonPRB05} or in single wall carbon
nanotubes with top gate electrodes.
\cite{graberPRB06,jorgensenAPL06, sapmazNL06,graber06} In such
devices by changing the respective gate voltages one can tune the
charge on each dot separately and also change the strength of the
coupling $t$ between the two dots. Furthermore, we also note that
in DQDs the on-level interaction $U$ is usually larger than the
inter-dot interaction $U^\prime$. In the case where the DQD is
doubly occupied and $U
> U^\prime$, the system can be in four different states
$\ket{\chi}$, namely
$\ket{\uparrow\uparrow}=\ket{\uparrow}\ket{\uparrow}$,
$\ket{\uparrow\downarrow}=\ket{\uparrow}\ket{\downarrow}$,
$\ket{\downarrow\uparrow}=\ket{\downarrow}\ket{\uparrow}$,
$\ket{\downarrow\downarrow}=\ket{\downarrow}\ket{\downarrow}$,
where the first (second) ket corresponds to the left (right) dot.
The occupation of the other two-particle states $\ket{\rm d
0}=\ket{\uparrow\downarrow}\ket{0}$ and $\ket{0\rm d} =
\ket{0}\ket{\uparrow\downarrow}$ is suppressed due to large
on-level interaction on the dots.

In the Coulomb blockade the charge fluctuations are suppressed and
the system is in a well-defined charge state. As a consequence,
all tunneling processes leading to a change of the DQD charge
state are exponentially suppressed. The current can thus flow due
to higher-order tunneling processes (cotunneling) through virtual
states in the double quantum dot.
\cite{weymannPRB05,nazarov,kang,averin92} The lowest-order
processes which give a dominant contribution to electric current
flowing through the DQD structure in the case of Coulomb blockade
are the third-order tunneling processes. Generally, the rate for
an $n$th-order ($n\geq 2$) tunneling from lead $j$ to lead
$j^\prime$ associated with a change of the double dot state from
$\chi$ into $\chi^\prime$ is given by \cite{averin92}
\begin{widetext}
\begin{equation}\label{Eq:rate}
  {\gamma^{(n)}}_{\rm jj^\prime}^{\chi\rightarrow\chi^\prime} =
    \frac{2\pi}{\hbar}\left|
    \sum_{v_1,v_2,\dots,v_{n-1}}
    \frac{
     \bra{\Phi_{j^\prime}^{\chi^\prime}}H_{\rm T}\ket{\Phi_{v_1}}
     \bra{\Phi_{v_1}}H_{\rm T}\ket{\Phi_{v_2}}\times \dots \times
     \bra{\Phi_{v_{n-1}}} H_{\rm T} \ket{\Phi_{j}^\chi}}
     {(\varepsilon_i-\varepsilon_{v_1})(\varepsilon_i-\varepsilon_{v_2})
     \times\dots\times(\varepsilon_i-\varepsilon_{v_{n-1}})}
     \right|^2 \delta (\varepsilon_i-\varepsilon_f)\,,
\end{equation}
\end{widetext}
where $\varepsilon_i$ and $\varepsilon_f$ denote the energies of
initial and final states and $\ket{\Phi_j^\chi}$ is the state of
the system with an electron in the lead $j$ and the double dot in
state $\ket{\chi}$, while $\ket{\Phi_{v_n}}$ denotes a virtual
state of the DQD system and $\varepsilon_{v_n}$ its energy. From
the above expression one can determine the third-order ($n=3$)
tunneling rates that give the main contribution to electric
current. We note that there are also tunneling events that do not
affect the DQD charge state but can have an influence on
transport. These are the second-order processes which take place
through a single tunnel barrier, either left or right. Such
single-barrier processes contribute to the electric current in an
indirect way, namely by changing the occupation probabilities and
this way the current. The rate of single-barrier second-order
cotunneling is given by Eq.~(\ref{Eq:rate}) for $n=2$ and
$j=j^\prime$. It is also worth noting that among different
higher-order tunneling events one can distinguish the {\it
elastic} (non-spin-flip) and {\it inelastic} (spin-flip)
processes. The former ones change the state of the double dot
$(\chi \neq \chi^\prime)$, while the latter ones do not $(\chi =
\chi^\prime)$.

\begin{figure}[b]
  \includegraphics[width=1\columnwidth]{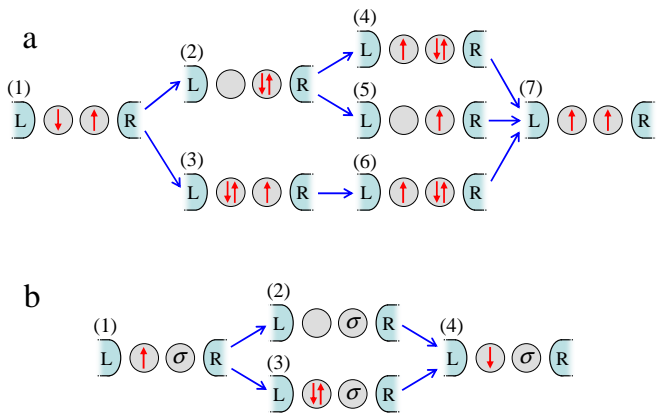}
  \caption{\label{Fig:2} (color online)
  Examples of possible tunneling processes through double quantum dot in the
  Coulomb blockade regime. The third-order process (a) from
  left to right lead, where
  $\ket{\uparrow\downarrow}$ is the initial state (1) and
  $\ket{\uparrow\uparrow}$ is the final state (7), takes place
  {\it via} five virtual states (2)-(6). This process contributes directly
  to the current flowing through the system. The second-order process
  through the left barrier (b), with $\ket{\uparrow\sigma}$
  ($\ket{\downarrow\sigma}$) being the initial (1) [finite (4)]
  state, where $\sigma=\uparrow,\downarrow$,
  affects the occupations of the double quantum dot.
  This process takes place {\it via} two virtual states (2)-(3).}
\end{figure}

Examples of possible processes in the case of the Coulomb blockade
regime are shown in Fig.~\ref{Fig:2}. The upper part of the figure
presents a third-order process from the left to right lead which
contributes to electric current. This is an inelastic process
which leads to a change of the double dot state from
$\ket{\downarrow\uparrow}$ to $\ket{\uparrow\uparrow}$. It takes
place through five virtual states, as sketched in
Fig.~\ref{Fig:2}a. On the other hand, the bottom part of
Fig.~\ref{Fig:2} displays a single-barrier second-order process,
occurring {\it via} two virtual states. This process does not
contribute to electric current but affects the DQD occupation
probabilities. The process shown in Fig.~\ref{Fig:2}b takes place
through the left barrier and changes the double dot state from
$\ket{\uparrow\sigma}$ into $\ket{\downarrow\sigma}$, with
$\sigma=\uparrow,\downarrow$. To make the discussion more
transparent, in Appendix we present the explicit formulas for the
rates corresponding to the two processes shown in
Fig.~\ref{Fig:2}.

By calculating all the second-order single-barrier and third-order
rates one can determine the occupation probabilities from the
following stationary master equation
\begin{eqnarray}
  0 &=& \sum_{jj^\prime}\sum_{\chi^\prime} \left[-\left(
  {\gamma^{(2)}}_{jj}^{\chi\rightarrow\chi^\prime}+
  {\gamma^{(3)}}_{jj^\prime}^{\chi\rightarrow\chi^\prime}\right) P_{\chi}
  \right.\nonumber \\
  &+& \left. \left( {\gamma^{(2)}}_{jj}^{\chi^\prime\rightarrow\chi}
  +{\gamma^{(3)}}_{jj^\prime}^{\chi^\prime\rightarrow\chi}\right) P_{\chi^\prime}
  \right] \,,
\end{eqnarray}
where $P_\chi$ denotes the probability for the double dot to be in
state $\ket{\chi}$. The occupations are fully determined with the
aid of the normalization condition, $\sum_{\chi}P_\chi = 1$. The
third-order current flowing through the system from the left to
right lead is then given by
\begin{equation}
  I = e \sum_{\chi\chi^\prime} P_\chi \left[
  {\gamma^{(3)}}^{\chi \rightarrow \chi^\prime}_{\rm LR} -
  {\gamma^{(3)}}^{\chi \rightarrow \chi^\prime}_{\rm RL} \right]\,.
\end{equation}

We note that generally the use of the master equation approach may
lead to wrong results in the regime where the level
renormalization effects or the effects due to exchange field
become important, i.e. close to resonance or for noncollinear
magnetic configurations.
\cite{weymannPRB05,weymannPRB07,koenigPRL03,braunPRB04} However,
in the following we consider only the case of deep Coulomb
blockade and collinear configurations, which justifies the
employed approach. \cite{weymannPRBBR05,weymannPRB06,weymannEPL06}

\section{Results and discussion}

We first present the results for cotunneling through double
quantum dots coupled to nonmagnetic leads, $p=0$. Next, we analyze
the case when the leads are ferromagnetic ($p>0$) and the system
can be either in parallel or antiparallel magnetic configuration.
The current flowing through the system depends then on the
magnetic configuration giving rise to tunnel magnetoresistance.
The TMR is qualitatively defined as
\cite{julliere75,barnas98,rudzinski01} ${\rm TMR} = I_{\rm
P}/I_{\rm AP} - 1$, where $I_{\rm P}$ ($I_{\rm AP}$) is the
current flowing through the system in the parallel (antiparallel)
magnetic configuration.

\subsection{DQD coupled to nonmagnetic leads}

\begin{figure}[t]
  \includegraphics[width=0.75\columnwidth]{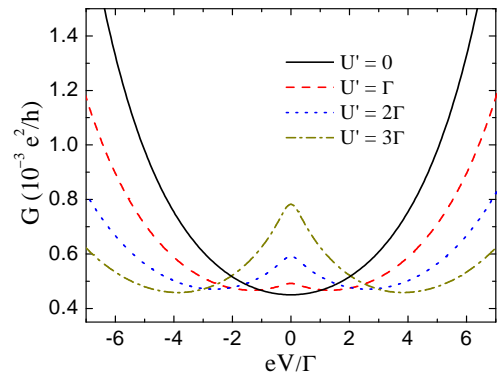}
  \caption{\label{Fig:3} (color online)
  Differential conductance as a function of the bias voltage
  for different inter-dot interaction parameter $U^\prime$,
  as indicated in the figure. The other parameters are
  $k_{\rm B} T = 0.1 \Gamma$,
  $\varepsilon=-10\Gamma$, $U=20\Gamma$, $t=\Gamma$,
  and $p=0$.}
\end{figure}

The differential conductance $G$ of the DQD coupled to nonmagnetic
leads as a function of the bias voltage for several values of the
inter-dot interaction parameter $U^\prime$ is shown in
Fig.~\ref{Fig:3}. In the case of negligible $U^\prime$, the
differential conductance exhibits a smooth parabolic dependence on
the bias voltage. However, for a finite inter-dot correlation,
there is a maximum in $G$ at zero bias. As one can see in the
figure, the magnitude of this maximum increases with increasing
$U^\prime$. Furthermore, when increasing $U^\prime$, the minimum
of $G$ at $V=0$ splits into two minima, separated by the zero-bias
peak.

In order to understand the mechanism leading to such behavior, we
note that in the spinless case, $p=0$, the occupation
probabilities do not depend on the applied voltage and are equal
to $1/4$, i.e. each of the four DQD states,
$\ket{\uparrow\uparrow}$, $\ket{\uparrow\downarrow}$,
$\ket{\downarrow\uparrow}$, $\ket{\downarrow\downarrow}$, is
equally occupied. Furthermore, the single-barrier second-order
processes which provide a channel for spin relaxation in the dots
do not have any influence on transport either. As a consequence,
the zero-bias maximum results from intrinsic dependence of
third-order tunneling rates on the value of the inter-dot
correlation parameter $U^\prime$.

The electric current flows through the DQD system due to
third-order processes which involve correlated tunneling through
virtual states of the system. More specifically, these virtual
states include the single-particle states, $\ket{\sigma 0}$ and
$\ket{0 \sigma}$, the two-particle states, $\ket{0 \rm d}$ and
$\ket{\rm d 0}$, and the three-particle states, $\ket{\sigma \rm
d}$ and $\ket{\rm d \sigma}$. In equilibrium, the energy of these
virtual states is respectively given by $\varepsilon_1 =
\varepsilon$, $\varepsilon_2 = 2\varepsilon+U$, and $\varepsilon_3
= 3\varepsilon+U+2U^\prime$. On the other hand, the energy of the
initial state is $\varepsilon_i = 2\varepsilon+U^\prime$.
Consequently, the resolvents that determine the rates, see Eq.
(\ref{Eq:rate}), are given by $(\varepsilon_i -
\varepsilon_1)^{-1} = (\varepsilon+U^\prime)^{-1}$,
$(\varepsilon_i - \varepsilon_2)^{-1} = (U^\prime-U)^{-1}$,
$(\varepsilon_i - \varepsilon_3)^{-1} = (- \varepsilon -U
-U^\prime)^{-1}$. The third-order tunneling processes take place
{\it via} two consecutive virtual states. Thus, the rate is
proportional to the product of two resolvents, depending on
virtual states being involved in a process. Generally, one can
distinguish four different contributions -- the first one involves
two single-particle states, $(\varepsilon+U^\prime)^{-2}$, the
second one involves one single-particle and one two-particle
state, $[(\varepsilon+U^\prime)(U^\prime-U)]^{-1}$, the third
contribution comes from one two-particle and one three-particle
state, $[(U-U^\prime)(\varepsilon+U+U^\prime)]^{-1}$, and the last
one involves two three-particle states,
$(\varepsilon+U+U^\prime)^{-2}$. After a crude estimation, one can
see from the above formulas that by increasing $U^\prime$, the
contribution coming from the first two resolvents is increased,
the third one is roughly constant, while that of the last
resolvent is decreased. This generally leads to an asymmetry of
cotunneling through different virtual states. Such asymmetry gives
rise to an enhancement of the conductance through the system by
increasing the rate of processes occurring {\it via} one-particle
and two-particle DQD states. As a result, with increasing
$U^\prime$, a maximum develops in the differential conductance at
the zero bias, see Fig.~\ref{Fig:3}. On the other hand, for a
given value of $U^\prime$, the differential conductance decreases
with increasing the bias voltage and reaches a minimum at $|eV|
\approx 2 U^\prime$. At this bias voltage the effect of finite
inter-dot Coulomb interaction is compensated by the transport
voltage, and the differential conductance reaches minimum, which
is present on both sides of the zero-bias anomaly.
Moreover, although the position of the two minima depends on the
inter-dot correlation, its value is rather independent of
$U^\prime$, see Fig.~\ref{Fig:3}.

When considering the case of linear response, zero temperature and
negligible inter-dot correlation, the minimum value of the
differential conductance can be approximated by the following
formula
\begin{equation}
   G = \frac{e^2t^2\Gamma^2}{2h}\left[
   \frac{1}{\varepsilon^2}
   -\frac{1}{\varepsilon(\varepsilon+U)}
   +\frac{1}{(\varepsilon+U)^2}
   \right]^2 \,.
\end{equation}
For the parameters assumed to calculate Fig.~\ref{Fig:3}, from the
above formula one finds, $G=0.45\times 10^{-3}\; {\rm e^2/h}$,
which is in good agreement with numerical results.

\subsection{DQD coupled to ferromagnetic leads}

If the leads are ferromagnetic ($p\neq 0$), the single-barrier
second-order processes start to influence transport by affecting
the DQD occupation probabilities. Transport characteristics are
then a result of the interplay between processes driving the
current and processes leading to spin relaxation in the dots.
First, we note that the rate of single-barrier processes is
proportional to temperature, while that of third-order processes
depends on the applied bias voltage, see Eqs.~(\ref{Eq:rate3}) and
(\ref{Eq:rate2}). This will give rise to interesting phenomena,
depending on the relative ratio of the second-order and
third-order processes, as will be discussed in the following.

\begin{figure}[t]
  \includegraphics[height=9cm]{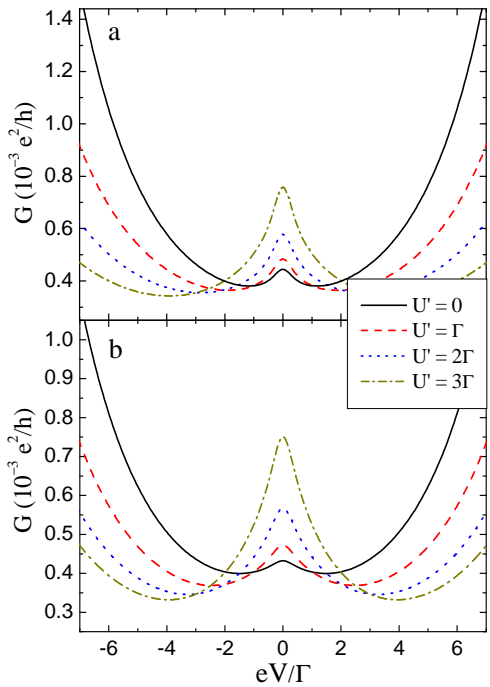}
  \caption{\label{Fig:4} (color online)
  Bias dependence of the differential conductance in
  the parallel (a) and antiparallel (b)
  magnetic configurations
  for different inter-dot interaction parameter $U^\prime$,
  as indicated. The other parameters are
  $k_{\rm B} T = 0.1 \Gamma$,
  $\varepsilon=-10\Gamma$, $U=20\Gamma$, $t=\Gamma$,
  and $p = 0.5$.}
\end{figure}

In Fig.~\ref{Fig:4} we show the bias dependence of the
differential conductance for the parallel and antiparallel
magnetic configurations of the system for several values of the
inter-dot interaction parameter $U^\prime$. First of all, it can
be seen that the value of $G$ at the zero bias increases with
increasing the inter-dot correlation. This is a general feature
which is present in both magnetic configurations of the system and
gives rise to the zero-bias maximum, see Fig.~\ref{Fig:4}a and b.
The mechanism leading to such behavior was already discussed in
the nonmagnetic case, i.e. a finite value of $U^\prime$ results in
increased cotunneling through one-particle and two-particle
virtual states, which in turn leads to an enhancement of the
differential conductance at the zero bias.

Another feature visible in the case of ferromagnetic leads is that
even for negligible $U^\prime$ there is a small maximum in $G$ at
the zero bias, irrespective of magnetic configuration of the
system. This maximum bears a resemblance to the zero-bias anomaly
found in the case of single quantum dots. \cite{weymannPRBBR05}
However, in single quantum dots the maximum is present only in the
antiparallel configuration, while in the case of double quantum
dots, interestingly, the zero-bias peak is present in both
magnetic configurations, see Fig.~\ref{Fig:4}a and b. In order to
understand this behavior we note that when there is a finite bias
voltage applied to the system, a nonequilibrium spin accumulation
can build up in the DQD. More precisely, for positive bias voltage
in the parallel configuration one observes unequal occupation of
singlet states, $P_{\ket{\downarrow\uparrow}} >
P_{\ket{\uparrow\downarrow}}$, while triplets are roughly equally
occupied (no spin accumulation), $P_{\ket{\uparrow\uparrow}}
\approx P_{\ket{\downarrow\downarrow}}$. On the other hand, in the
antiparallel configuration there is unequal occupation of triplet
states (spin accumulation), $P_{\ket{\downarrow\downarrow}} >
P_{\ket{\uparrow\uparrow}}$, whereas singlets are equally
occupied, $P_{\ket{\downarrow\uparrow}} \approx
P_{\ket{\uparrow\downarrow}}$. It is further interesting to
realize that for positive bias voltage main contribution to the
current comes from third-order tunneling processes having the
initial state $\ket{\uparrow\downarrow}$ for the parallel and
$\ket{\uparrow\uparrow}$ for the antiparallel magnetic
configuration. Thus, with increasing the bias voltage ($V>0$), the
contribution coming from those processes is decreased, leading to
a decreased conductance. As a consequence, one observes a maximum
at the zero bias even in the case of $U^\prime = 0$, see
Fig.~\ref{Fig:4}.

The zero-bias maximum in differential conductance is therefore a
result of superposition of two different effects. The first one
concerns the asymmetry of cotunneling through virtual states,
which is induced by a finite value of the inter-dot Coulomb
interaction. Whereas the second one is associated with unequal
occupation of the corresponding DQD states, which results from
spin-dependent tunneling rates.

\begin{figure}[t]
  \includegraphics[height=9cm]{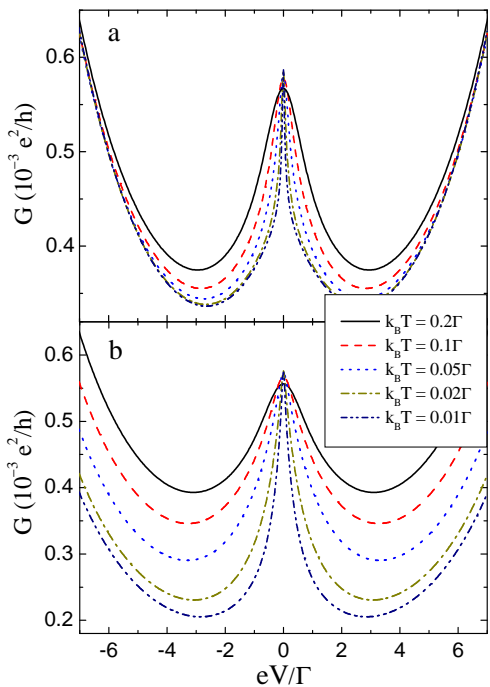}
  \caption{\label{Fig:5} (color online)
  Bias dependence of the differential conductance in
  the parallel (a) and antiparallel (b)
  magnetic configurations
  for different temperatures and for $U^\prime=2\Gamma$.
  The other parameters are the same as in Fig.~\ref{Fig:4}.}
\end{figure}

When the DQD is coupled to ferromagnetic leads, an important role
is played by the single-barrier second-order processes -- they do
not contribute to the current, but lead to the spin relaxation in
the double dot system. In order to gain more intuitive
understanding of the discussed phenomena, in the following we
present a crude quantitative analysis of the processes determining
transport behavior. When considering the low temperature limit and
assuming $U=-2\varepsilon$, $U^\prime=0$, the rate of
single-barrier second-order processes can be approximated
by\cite{weymannPRB06}
\begin{equation}\label{Eq:rateapprox2}
  {\gamma^{(2)}}_{jj}^{\ket{\sigma \chi}\rightarrow
  \ket{\bar{\sigma} \chi}} \approx
  \frac{4k_{\rm B}T \Gamma^2}{h\varepsilon^2} \,.
\end{equation}
On the other hand, we note that generally the fastest third-order
processes are the ones leading to the change of the dot state from
$\ket{\sigma\bar{\sigma}}$ into $\ket{\bar{\sigma}\sigma}$. With
the same assumptions as made above, one can approximate the rate
of such processes by the following formula
\begin{equation}\label{Eq:rateapprox3}
  {\gamma^{(3)}}_{jj^\prime}^{\ket{\sigma\bar{\sigma}}\rightarrow
  \ket{\bar{\sigma}\sigma}} \approx
  \frac{16|eV|t^2\Gamma^2}{h\varepsilon^4}\,.
\end{equation}
The above expressions show explicitly that the relative ratio of
both processes depends on the internal system parameters as well
as the temperature and applied bias voltage. Furthermore, one can
now roughly estimate the bias voltage at which the corresponding
second-order and third-order processes become comparable, it is
given by
\begin{equation}\label{Eq:biasapprox}
  |eV| \approx \frac{k_{\rm B}T \varepsilon^2}{4t^2} \,.
\end{equation}
This formula will be helpful in discussing the temperature
dependence of transport characteristics.

The influence of temperature on the bias dependence of
differential conductance in both magnetic configurations is shown
in Fig.~\ref{Fig:5}. One can see that with increasing thermal
energy, the width of the zero-bias peak is increased, while the
maximum value of $G$ for $V=0$ stays rather unchanged. This is due
to the fact that by raising the temperature, one increases the
role of single-barrier second-order processes, see
Eq.~(\ref{Eq:rateapprox2}), giving rise to faster spin relaxation.
Spin relaxation in turn leads to a decrease in the spin
accumulation induced in the system. \cite{weymannPRB06} Therefore,
the temperature effects on the differential conductance are more
visible in the antiparallel configuration than in the parallel
one. By decreasing $T$, the relative role of second-order
processes is decreased, which leads to larger spin accumulation,
$P_{\ket{\downarrow\downarrow}}
> P_{\ket{\uparrow\uparrow}}$. This in turn gives rise to an
increased and more robust drop of the differential conductance
with the bias voltage, see for example the curves for $k_{\rm
B}T=0.2\Gamma$ and $k_{\rm B}T=0.01\Gamma$ in Fig.~\ref{Fig:5}. As
a consequence, with decreasing temperature, the value of the
differential conductance at the minimum is decreased and the width
of the zero-bias peak becomes smaller -- the two minima in $G$
appear at smaller bias voltage. This is due to the fact that the
relative ratio of the second-order and third-order processes
changes with changing $T$ and, consequently, the bias voltage at
which the rates of these two processes are comparable is changed,
see Eq.~(\ref{Eq:biasapprox}). The dependence of the differential
conductance on temperature in the parallel configuration is less
pronounced than in the antiparallel configuration because for the
parallel configuration the single-barrier spin-flip processes only
slightly affect the DQD occupations. This results from the fact
that in the parallel configuration there is a left-right symmetry
between the couplings to the spin-majority and spin-minority
electron subbands. \cite{weymannPRBBR05}

We also note that in the spinless case discussed in previous
subsection the single-barrier second-order processes do not affect
transport in any way, and the occupations of all DQD states are
equal. Therefore, the differential conductance only slightly
depends on temperature.

\begin{figure}[t]
  \includegraphics[width=0.75\columnwidth]{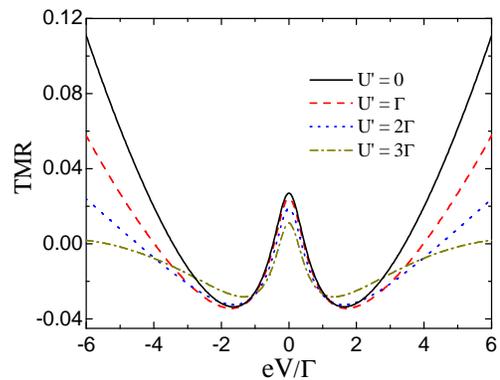}
  \caption{\label{Fig:6} (color online)
  Bias dependence of the TMR
  for different inter-dot interaction parameter $U^\prime$
  and for $k_{\rm B} T = 0.1 \Gamma$.
  The other parameters are the same as in Fig.~\ref{Fig:4}.}
\end{figure}

In Fig.~\ref{Fig:6} we present the TMR as a function of the bias
voltage for several values of the inter-dot correlation parameter.
First of all, it can be seen that for low bias voltages tunnel
magnetoresistance is only slightly affected by the inter-dot
interaction. This is due to the fact that the asymmetry in
tunneling through virtual states induced by finite value of
$U^\prime$ changes transport characteristics in both magnetic
configurations in a similar way, see Fig.~\ref{Fig:4}. As a
consequence, the TMR, which reflects the difference between the
parallel and antiparallel magnetic configuration, is roughly
independent of the value of inter-dot correlation.

Another interesting feature visible in Fig.~\ref{Fig:6} is the
sign change of the TMR -- with increasing the bias voltage, tunnel
magnetoresistance decreases from a maximum at the zero bias to a
minimum, at which TMR changes sign and becomes negative. At this
bias voltage conductance in the parallel configuration is smaller
than in the antiparallel configuration. This seemingly
counterintuitive fact can be understood when one takes into
account the effect of second-order processes giving rise to spin
relaxation. As already mentioned, in the parallel configuration
one finds, $P_{\ket{\uparrow\downarrow}} \neq
P_{\ket{\downarrow\uparrow}}$, while in the antiparallel
configuration one has, $P_{\ket{\uparrow\uparrow}} \neq
P_{\ket{\downarrow\downarrow}}$. Spin relaxation processes
decrease the spin accumulation in the antiparallel configuration,
which leads to an enhancement of the differential conductance, see
Fig.~\ref{Fig:5}b. On the other hand, in the parallel
configuration the DQD occupations only slightly depend on
second-order processes. As a consequence, if the spin relaxation
processes are sufficiently fast, ${\gamma^{(2)}}_{jj}^{\ket{\sigma
\chi}\rightarrow \ket{\bar{\sigma} \chi}} \gtrsim
{\gamma^{(3)}}_{jj^\prime}^{\ket{\sigma\bar{\sigma}}\rightarrow
\ket{\bar{\sigma}\sigma}}$, one observes negative TMR effect.

\begin{figure}[t]
  \includegraphics[width=0.75\columnwidth]{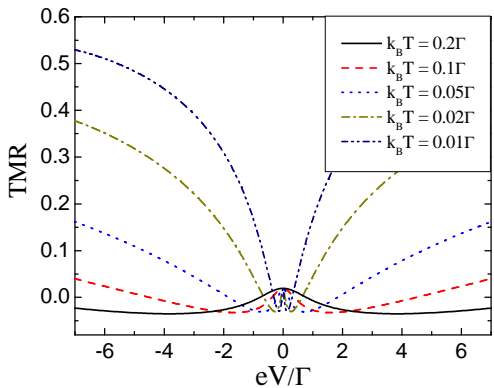}
  \caption{\label{Fig:7} (color online)
  Tunnel magnetoresistance as a function of the bias voltage
  for different temperatures and for $U^\prime=2\Gamma$.
  The other parameters are the same as in Fig.~\ref{Fig:4}.}
\end{figure}

In Fig.~\ref{Fig:7} we display the TMR effect as a function of the
bias voltage for different temperatures. First of all, one can see
that TMR exhibits a nontrivial dependence on temperature. This is
because by changing $T$, one effectively changes the amount of
processes leading to spin relaxation which affect spin
accumulation and, thus, conductance in the antiparallel
configuration. For low temperatures, second-order processes are
suppressed and TMR becomes positive in the whole range of the bias
voltage with a minimum at the zero bias, see the curve for $k_{\rm
B}T = 0.01\Gamma$ in Fig.~\ref{Fig:7}. On the other hand, for
higher temperatures the rate of single-barrier second-order
processes is increased, which gives rise to two minima in the TMR
separated by the zero-bias maximum, see Figs.~\ref{Fig:6} and
\ref{Fig:7}. Moreover, at these minima TMR changes sign and
becomes negative. We note that the negative TMR was also observed
in single quantum dots in the limit of fast spin relaxation in the
dot. \cite{weymannPRB06}

Finally, we present analytical formulas approximating tunnel
magnetoresistance in the most characteristic transport regimes.
For $|eV| \gg k_{\rm B}T$, the TMR can be expressed as
\begin{equation}
  {\rm TMR} =
  \frac{2p^2}{1-p^2}\frac{53-3p^2(1+7p^2-p^4)}{(5+3p^2)(3-p^2)^2}\,,
\end{equation}
where we have assumed the symmetric Anderson model for each dot,
$U=-2\varepsilon$, and $U^\prime=0$. This formula approximates the
TMR in the zero temperature limit, i.e. in the absence of
second-order processes. On the other hand, the linear response TMR
calculated with the same assumptions can be approximated by
\begin{equation}
  {\rm TMR} = \frac{2p^2}{1-p^2}\frac{13+3p^4}{(9+p^2)(5+3p^2)}
  \,.
\end{equation}

\section{Concluding remarks}

We have considered cotunneling transport through double quantum
dots in series weakly coupled to ferromagnetic leads. In the
Coulomb blockade regime the current flows through the system due
to third-order tunneling processes. We have also taken into
account the single-barrier second-order processes which do not
contribute to the current but affect the DQD occupation
probabilities.

We have shown that the differential conductance exhibits a maximum
at the zero bias, irrespective of magnetic configuration of the
system. This anomalous behavior results from the superposition of
two different effects. The first effect is associated with
asymmetry of cotunneling through different virtual states which
can be induced by the inter-dot Coulomb interaction. The second
mechanism results from the interplay of single-barrier
second-order processes leading to spin relaxation and the
third-order tunneling processes contributing to the current. The
first mechanism does not depend on the value of spin polarization
of the leads, the second one, on the contrary, results from the
spin dependency of tunneling rates.

We have also analyzed the temperature dependence of transport
characteristics. By changing thermal energy, one effectively
changes the rate of the second-order processes, i.e. the amount of
spin relaxation processes. We have shown that the width of the
zero-bias maximum in the differential conductance increases with
increasing temperature. This effect is most visible in the
antiparallel configuration, which is due to the fact that in the
antiparallel configuration spin relaxation decreases the spin
accumulation induced in the DQD system, while occupations in the
parallel configuration only slightly depend on the spin
relaxation.

Furthermore, we have also shown that TMR exhibits a nontrivial
dependence on temperature. For low temperatures, the TMR exhibits
a minimum at the zero bias. However, for higher temperatures this
minimum splits into two minima separated by a maximum at the zero
bias. At the these minima tunnel magnetoresistance changes sign
and becomes negative.


\begin{acknowledgments}

We acknowledge discussions with J. Barna\'s. This work, as part of
the European Science Foundation EUROCORES Programme SPINTRA, was
supported by funds from the Ministry of Science and Higher
Education as a research project in years 2006-2009 and the EC
Sixth Framework Programme, under Contract N. ERAS-CT-2003-980409,
and the Foundation for Polish Science.

\end{acknowledgments}


\appendix

\section{Examples of cotunneling rates}

In the following we present the explicit formulas for the
third-order and second-order tunneling rates corresponding to
processes shown in Fig.~\ref{Fig:2}a and b. To determine the rate
${\gamma^{(3)}}^{\ket{\downarrow\uparrow}\rightarrow
\ket{\uparrow\uparrow}}_{\rm LR}$ one needs to find the initial
and final energies of the whole process, as well as the energies
of the virtual states, as sketched in Fig.~\ref{Fig:2}a. Then, by
calculating the respective energy differences and plugging them
into Eq.~(\ref{Eq:rate}), one finds
\begin{eqnarray}\label{Eq:rate3}
   {\gamma^{(3)}}^{\ket{\downarrow\uparrow}\rightarrow
   \ket{\uparrow\uparrow}}_{\rm LR} =
   \frac{\Gamma_{\rm L}^\uparrow \Gamma_{\rm R}^\downarrow t^2}{h}
   \int d\omega f^+(\omega)f^-(\omega+\mu_{\rm L}-\mu_{\rm R}) \nonumber\\
   \times \left[\frac{1}{(\omega+\mu_{\rm
   L}-\varepsilon-U-U^\prime)^2} -
   \frac{1}{U-U^\prime}\right.\nonumber\\
   \left.\times\left(
   \frac{1}{\omega+\mu_{\rm L}-\varepsilon-U-U^\prime}
   -\frac{1}{\omega+\mu_{\rm L}-\varepsilon-U^\prime}
   \right)
   \right]^2 \,,
\end{eqnarray}
where $f^+$ is the Fermi function and $f^- = 1-f^+$. On the other
hand, the single-barrier second-order rate for the process shown
in Fig.~\ref{Fig:2}b can be found in a similar way. This rate is
given by
\begin{eqnarray}\label{Eq:rate2}
   {\gamma^{(2)}}^{\ket{\uparrow\sigma}\rightarrow
   \ket{\downarrow\sigma}}_{\rm LL} =
   \frac{\Gamma_{\rm L}^\uparrow \Gamma_{\rm L}^\downarrow}{h}
   \int d\omega f^+(\omega)f^-(\omega) \nonumber\\
   \times \left[\frac{1}{\omega+\mu_{\rm
   L}-\varepsilon-U-U^\prime}
   -\frac{1}{\omega+\mu_{\rm L}-\varepsilon-U^\prime}
   \right]^2 \,.
\end{eqnarray}
We note that in the simplest approximation \cite{weymannEPJ05} for
the Coulomb blockade regime one can pull out the resolvents in
front of the integrals. Then, one arrives at the following
integral, $\int d\omega f^+(\omega)f^-(\omega+\xi)$, where
$\xi=\mu_{\rm L}-\mu_{\rm R}$ or $\xi=0$, correspondingly, which
can be easily calculated. \cite{averin92} As a consequence, one
can see that the rate of single-barrier second-order processes is
proportional to temperature $T$, whereas that of the third-order
processes depends on the bias voltage $V$.


\end{document}